\documentstyle[11pt,fleqn]{article}
\newcommand{\preprint}[1]{\hfill{\sl preprint - #1}\par\bigskip\par\rm}
\def\titolo{\par\bigskip\begin{center}\bf\LARGE}
\def\endtitolo{\end{center}\par\bigskip\par\rm\normalsize}
\def\instit{\begin{center}\large}
\def\endinstit{\end{center}\rm\normalsize}
\def\references{\begin{thebibliography}{10}}
\def\endreferences{\end{thebibliography}\end{document}}
\newcommand{\dip}{\smallskip Dipartimento di Fisica,
                                Universit\`a di Trento, Italia}
\newcommand{\infn}{\smallskip Istituto Nazionale di Fisica Nucleare,\\
                                 Gruppo Collegato di Trento, Italia}
\newcommand{\dinfn}{\dip\\ and \infn}
\newcommand{\btit}{\begin{titolo}}
\newcommand{\etit}{\end{titolo}}

\newcommand{\Idinfn}{\begin{instit}\dinfn\end{instit}}

\renewcommand{\author}[1]{\begin{center}\Large #1\end{center}}
\renewcommand{\date}[1]{\par\bigskip\par\sl\hfill #1\par\medskip\par\rm}
\newcommand{\email}[1]{e-mail: \sl #1@science.unitn.it,
                               #1@itncisca.bitnet, 37953::#1\rm}
\newcommand{\femail}[1]{\footnote{\email{#1}}}
\newcommand{\pacs}[1]{\smallskip\noindent{\sl PACS number(s):
                       \hspace{0.3cm}#1}\par\bigskip\rm}
\newcommand{\babs}{\hrule\par\begin{description}\item{Abstract: }\it}
\newcommand{\eabs}{\par\end{description}\hrule\par\medskip\rm}
\newcommand{\ack}[1]{\par\section*{Acknowledgments} #1}
\renewcommand{\vec}[1]{{\bf #1}}       
\newcommand{\M}{{\cal M}}              
\newcommand{\hs}{\qquad\qquad}         
\newcommand{\nn}{\nonumber}            
\newcommand{\beq}{\begin{eqnarray}}    
\newcommand{\eeq}{\end{eqnarray}}      
\newcommand{\beqn}{\begin{eqnarray}}   
\newcommand{\eeqn}{\end{eqnarray}}     
\newcommand{\ap}{\left.}               
\newcommand{\at}{\left(}               
\newcommand{\aq}{\left[}               
\newcommand{\ag}{\left\{}              
\newcommand{\cp}{\right.}              
\newcommand{\ct}{\right)}              
\newcommand{\cq}{\right]}              
\newcommand{\cg}{\right\}}             
\newcommand{\ii}{\infty}                         
\newcommand{\X}{\times\,}                        
\newcommand{\fr}[2]{\mbox{$\frac{#1}{#2}$}}      
\newcommand{\Tr}{\,\mbox{Tr}\,}                  
\newcommand{\PP}{\,\mbox{PP}\,}                  
\newcommand{\Res}{\,\mbox{Res}\,}                
\renewcommand{\Re}{\,\mbox{Re}\,}                
\newcommand{\lap}{\Delta}                        

\newcommand{\be}{\beta}

\newcommand{\de}{\delta}
\newcommand{\ep}{\varepsilon}
\newcommand{\ze}{\zeta}

\newcommand{\la}{\lambda}

\newcommand{\si}{\sigma}

\newcommand{\ph}{\varphi}

\newcommand{\te}{\vartheta}

\newcommand{\Ga}{\Gamma}

\begin{document}

\preprint{UTF 342}
\btit
One-loop Quantum Corrections to the Entropy\\
for a 4-dimensional Eternal Black Hole
\etit
\author{Guido Cognola\femail{cognola}, Luciano Vanzo\femail{vanzo} and
Sergio Zerbini\femail{zerbini}}
\Idinfn

\date{January - 1995}

\babs
The first quantum corrections to the free energy for an eternal
4-dimensional black hole is investigated at one-loop level, in the
large mass limit of the black hole, making use of the conformal
techniques  related to the optical metric. The quadratic and
logarithmic divergences as well as a finite part associated with the
first quantum correction to the entropy are obtained at a generic
temperature. It is argued that, at the Hawking temperature, the
horizon divergences of the internal energy should cancel. Some
comments on the divergences of the entropy are also presented. \eabs


Recently, several issues like the interpretation of the
Bekenstein-Hawking classical formula for the black hole entropy, the
loss of information paradox and the validity of the area law have been
discussed in the literature (see, for example, the review
\cite{beke94u-15}). There have also been some attemps to compute
semiclassically the first quantum corrections to the
Bekenstein-Hawking classical entropy  $4\pi G M^2$
\cite{beke73-7-2333,hawk75-43-199}. However, so far all the
evaluations have been plagued by the appearance of  divergences
\cite{thoo85-256-727,bomb86-34-373,suss94-50-2700,dowk94-11-55,solo94u-246,furs94u-32,empa94u-5}
present also in the related "entanglement or geometric entropy"
\cite{sred93-71-666,call94-333-55,kaba94-329-46}.

In the black hole case, the physical origin of these divergences can
be traced back to the equivalence principle
\cite{isra76-57-107,scia81-30-327,barb94-50-2712,dabh94u-68},
according to which, in a static space-time with canonical horizons, a
system in thermal equilibrium has a local Tolman temperature  given by
$T(x)=T/{\sqrt {-g_{00}(x)}}$, $T$ being the generic asymptotic
temperature. Since, roughly speaking, very near the horizon a static
space-time may be regarded as a Rindler-like space-time, one gets for
the Tolman temperature $T(\rho)=T/\rho$,  $\rho$ being the distance
from the horizon. As a consequence, omitting the multiplicative
constant, the total entropy reads \beq S\sim\int d{\vec x}\int_\ep^\ii
T^3(\rho)\,d\rho= \frac{AT^3}{2\ep^2} \:,\label{1.3} \eeq where $A$ is
the area of the horizon and $\ep$ the horizon cutoff. These
considerations suggest that the use of the optical metric
$\bar{g}_{\mu\nu}=g_{\mu\nu}/|g_{00}|$, conformally related to the
original one, may provide an alternative and useful framework to
investigate these issues and one purpose of this paper is to implement
this idea, which has been proposed in Refs.
\cite{page82-25-1499,brow85-31-2514} and put recently forward also in
Refs.~\cite{frol94u-36,barv94u-36,empa94u-5,deal94u-27} for the cases
of black hole and Rindler space-times.

To start with, we consider a  scalar field on a 4-dimensional static
space-time defined by the metric (signature $-+++$) \beq
ds^2=g_{00}(\vec{x})(dx^0)^2+g_{ij}(\vec{x})dx^idx^j\:, \hs
\vec{x}=\{x^j\}\:,\hs i,j=1,...,3\:. \eeq

The one-loop partition function is given by (we perform the Wick
rotation $x_0=-i\tau$, thus all differential operators one is dealing
with will be elliptic) \begin{equation} Z=\int d[\phi]\,
\exp\at-\frac12\int\phi L_4 \phi d^4x\ct \:,\end{equation} where
$\phi$ is a scalar density of weight $-1/2$ and $L_4$ is a
Laplace-like operator on a 4-dimensional manifold. It has the form
\beq L_4=-\lap_4+m^2+\xi R \:.\eeq Here $\lap_4$ is the
Laplace-Beltrami operator, $m$ (the mass) and $\xi$ arbitrary
parameters and $R$ the scalar curvature of the manifold.

Now we recall the conformal transformation technique
\cite{dowk78-11-895,gusy87-46-1097,dowk88-38-3327,dowk89-327-267}.
This method is useful because it permits to compute all physical
quantities in an ultrastatic manifold (called the optical manifold
\cite{gibb78-358-467}) and, at the end of calculations, to transform
back them to a static one, with an arbitrary $g_{00}$. The ultrastatic
Euclidean metric $\bar{g}_{\mu\nu}$ is related to the static one by
the conformal transformation \beq
\bar{g}_{\mu\nu}(\vec{x})=e^{2\si(\vec{x})}g_{\mu\nu}(\vec{x}) \:,\eeq
with $\si(\vec{x})=-\frac{1}{2}\ln g_{00}$. In this manner,
$\bar{g}_{00}=1$ and $\bar{g}_{ij}=g_{ij}/g_{00}$ (Euclidean optical
metric).

For the one-loop partition function it is possible to show that \beq
\bar{Z}=J[g,\bar{g}]\,Z \:,\eeq where $J[g,\bar{g}]$ is the Jacobian
of the conformal transformation. Such a Jacobian can be explicitely
computed \cite{dowk89-327-267}, but here we shall need only its
structural form. Using $\zeta$-function regularization for the
determinant of the differential operator we get \beq \ln Z=\ln\bar
Z-\ln J[g,\bar g] =\frac{1}{2}\ze'(0|\bar L_4\ell^2)-\ln J[g,\bar{g}]
\:,\label{lnZ-Zbar}\eeq where $\ell$ is an arbitrary parameter
necessary to adjust the dimensions and $\ze'$ represents the
derivative with respect to $s$ of the function $\ze(s|\bar L_4\ell^2)$
related to the operator $\bar L_4$, which explicitly reads \beq \bar
L_4=e^{-\si}L_4e^{-\si}= -\partial_\tau^2-\bar\lap_3+\fr16\bar R
+e^{-2\si}\aq m^2+(\xi-\fr16)R\cq =-\partial_\tau^2+\bar L_3
\:.\label{aconf}\eeq

The same analysis can be easily extended to the finite temperature
case \cite{dowk88-38-3327}. In fact, we recall that for a scalar field
in thermal equilibrium at finite temperature $T=1/\be$ in an
ultrastatic space-time, the partition function $\bar{Z}_\be$ may be
obtained, within the path integral approach, simply by Wick rotation
$\tau=ix^0$ and imposing a $\be$ periodicity in $\tau$ for the field
$\bar\phi(\tau,x^i)$ \cite{bern74-9-3312,dola74-9-3320}. In this way,
in the one loop approximation one has \begin{equation}
\bar{Z}_\be=\int_{\bar\phi(\tau,x^i)=\bar\phi(\tau+\be,x^i)}
d[\bar\phi]\,\exp\at-\int_0^\be d\tau\int\bar\phi\bar
L_4\bar\phi\,d^3x\ct \:.\label{PF} \end{equation} Since the Euclidean
metric is $\tau$ independent, one obtains \beq \ln\bar
Z_\be&=&-\frac{\be}{2}\aq \PP\ze(-\fr12|\bar L_3)
+(2-2\ln2\ell)\Res\ze(-\fr12|\bar{L}_3)\cq\nn\\ &&\hs+\lim_{s\to0}
\frac{d}{ds}\frac{\be}{\sqrt{4\pi}\Ga(s)} \sum_{n=1}^\ii\int_0^\ii
t^{s-3/2}\,e^{-n^2\be^2/4t}\, \Tr e^{-t\bar{L}_3}\,dt
\label{lnZbeta}\:, \eeq where $\PP$ and $\Res$ stand for the principal
part and the residue of the function and one has to analytically
continue before taking the limit $s\to0$. As usual, in the definition
of $\zeta$-function the subtraction of possible zero-modes of the
corresponding operator is left understood. Of course, if the function
$\ze(s|\bar L_3)$ is finite for $s=-1/2$, the first term on the
right-hand side of the latter equation is just
$-\frac\be2\ze(-\frac12|\bar{L}_3)$. The latter formula is rigorously
valid for a compact manifold. In the paper we shall deal with a non
compact manifold, but nevertheless we shall make  a formal use of this
general formula, employing $\ze$-function associated with continuum
spectrum.

The free energy is related to the canonical partition function by
means of the equation \beq F_\be=E_v+\hat{F}_\be =-\frac{1}{\be}\ln
Z_\be =-\frac{1}{\be}\at\ln\bar Z_\be-\ln J[g,\bar g]\ct
\:,\label{FE}\eeq where $E_v$ is the vacuum energy while $\hat{F}_\be$
represents the temperature dependent part (statistical sum). It should
be noted that since we are considering a static space-time, the
quantity $\ln J[g,\bar g]$ depends linearly on $\be$ and, according to
Eq.~(\ref{FE}), it gives contributions only to the vacuum energy term
and not to entropy, which  may be defined by \beq S_\be=\be^2
\partial_\be F_\be \:, \label{entropy} \eeq and for the internal
energy we assume the well known thermodynamical relation \beq
U_\be=\frac{S_\be}{\be}+ F_\be \:.\label{u} \eeq

Let us apply this formalism to the case of a massless  scalar field in
the 4-dimensional Schwarzschild background with a generic
gravitational coupling $\xi R$, $\xi=1/6 $ being the conformal
coupling. Our aim is to compute the entropy of this field using the
latter formula. It may be considered as the prototype of the quantum
correction to the classical entropy of a black hole.

To begin with, we recall that the metric is \beq
ds^2=-\at1-\frac{r_H}r\ct\,(dx^0)^2+
\at1-\frac{r_H}r\ct^{-1}\,dr^2+r^2\,d\Omega_{2} \:,\label{bh} \eeq
where we are using polar coordinates, $r$ being the radial one and
$d\Omega_{2}$ the $2$-dimensional spherical unit metric. The horizon
radius is $r_H=2MG$, $M$ being the mass of the black hole and $G$ the
Newton constant. From now on, for the sake of convenience we put
$r_H=1$; in this way all quantities are dimensionless; the dimensions
will be easily restored at the end of calculations.

It may be convenient to redefine the Schwarzschild coordinates
$(x^0,r)$ by means \beq x'^0=\frac{x^0}2\:,\hs
\rho=2(r-1)^{\frac12}e^{(r-1)/2} \:.\label{ro} \eeq Thus, in the new
set of coordinates, we have \beq ds^2=-4\at1-\frac1r\ct\,(dx'^0)^2+
4\at1-\frac1r\ct\,\frac{d\rho^2}{\rho^2}+ r^2\,d\Omega_{2}
\:,\label{sus} \eeq where $r$ is implicitely defined by Eq.~(\ref{ro})
and has the expansion, valid near the horizon $r \sim 1$ or $\rho \sim
0$ \beq r\sim1+\frac{\rho^2}{4}-\frac{\rho^4}{16}+O(\rho^6)
\:.\label{456}\eeq The optical metric
$\bar{g}_{\mu\nu}=g_{\mu\nu}/(-g_{00})$, which is conformally related
to the previous one and appears as an ultrastatic metric, reads \beq
d\bar s^2=-(dx'^0)^2 +\frac1{\rho^2}\aq
d\rho^2+G(\rho)\,d\Omega_{2}\cq \label{OM}\eeq and has curvature given
by \beq \bar R=-\frac{6}{r^4} \sim -6+6\rho^2+O(\rho^4)
\:,\label{88}\eeq where \beq G(\rho)=r^3e^{(r-1)}\sim
1+\rho^2+O(\rho^4) \label{67}\:.\eeq In order to perform explicit
computations, we shall consider the large mass limit  of the black
hole and this leads to the approximated metric \beq d\bar
s^2\sim-(dx'^0)^2 +\frac{1}{\rho^2}\aq d\rho^2+d\Omega_{2}\cq
\label{lsus1} \:.\eeq This can be considered as an approximation of
the metric defined by Eq.~(\ref{OM}) in the sense that, near the
horizon $\rho=0$, the geodesics are essentially the same for both the
metrics \cite{unrh76-14-870}. Eq.~(\ref{lsus1}) defines a manifold
with curvature $\bar R=-6+2\rho^2$. Then, according to
Eq.~(\ref{aconf}), the relevant operator becomes \beq \bar
L_3=-\bar\lap_3-1+\at m^2+C \ct\rho^2 \label{barLN} \:,\eeq where
$\bar\lap_3$ is the related Laplace-Beltrami operator and the constant
$C$ acts as an effective mass and has been introduced in order to take
into account of the contribution to the curvature (at this order) of
the function $ G(\rho) $. Working within the first approximation,
defined by the metric (\ref{lsus1}), its value is $C=1/3$. It should
be noted the appearance of an effective "tachionic" mass $-1$, which
has important consequences on the structure of the $\zeta$-function
related to the operator $\bar L_3$. The invariant measure and the
Laplace-Beltrami operator read \beq d\bar
V&=&\rho^{-3}\,d\rho\,dV_{2}\:,\nn\\ \bar\lap_3&=&
\rho^2\partial_\rho^2-\rho\,\partial_\rho +\rho^2\lap_2 \label{da}
\:,\eeq where $\lap_2$ is the Laplace-Beltrami operator on the unitary
sphere $S^2$ and $dV_{2}$ its invariant measure.

In order to study the quantum properties of matter fields defined on
this ultrastatic manifold, it is sufficient to investigate the kernel
of the operator $e^{-t\bar{L}_3}$. The eigenfunctions of the operator
$-\lap_2+m^2+C$ are the spherical harmonics $Y_l^m(\te,\ph)$ and the
eigenvalues $\la_l^2=l(l+1)+m^2+C$. Let
$\Psi_{rlm}=\phi_{rl}(\rho)Y_l^m(\te,\ph)$ be the eingenfunctions of
$\bar L_3$ with eigenvalues $\la_r^2$. The differential equation which
determines the continuum spectrum turns out to be \beq
\aq\rho^2\,\partial_\rho^2 -\rho\,\partial_\rho -\rho^2
\la_l^2+\la_r^2+1 \cq\phi_{rl}(\rho)=0 \:.\label{Bessel} \eeq The only
solutions with the correct decay properties at infinity are the Bessel
functions of imaginary argument, with $\la_r^2=r^2\geq0$. Thus we have
\beq \phi_{rl}(\rho)=\rho K_{ir}(\rho\la_l) \:.\label{10} \eeq It
should be noted that the solutions are vanishing for $\rho=0$, even if
we are not assuming any boundary condition close to the horizon, as in
the "brick-wall" regularization of
Refs.~\cite{thoo85-256-727,suss94-50-2700}. As a consequence we have
to deal with a continuum spectrum.

For any suitable function $f(\bar L_3)$, we may write \beq <x|f(\bar
L_3)|x> =\int_0^\ii f(r^2)\sum_{lm}\mu_l(r)
Y_l^{*m}(\te,\ph)\phi^*_{rl}(\rho) Y_l^m(\te,\ph)\phi_{rl}(\rho)\,dr
\:,\label{hk} \eeq where $\mu_l(r)$ is the spectral measure associated
with the continuum spectrum. It is defined by means of equation \beq
\at\phi_{rl},\phi_{r'l}\ct= \frac{\de(r-r')}{\mu_l(r)} \:.\eeq Using
the asymptotic behaviour of the Mac Donald functions at the origin
\cite{grad80b} one can easily show that $\mu_l(r)$ does not depend on
$l$ and reads \beq \mu(r)\equiv\mu_l(r)=\frac2{\pi^2}\,r\sinh\pi r
\:.\label{v2} \eeq As a consequence, the heat kernel of $\bar L_3$ is
given by \beq K_t(x|\bar L_3)=\int_0^\ii \frac{2}{\pi^2}\,r\,\sinh\pi
r \sum_{l=0}^\ii\frac{(2l+1)}{4\pi}
\rho^{2}K_{ir}^2(\rho\la_l)\,e^{-tr^2}\,dr \:.\label{bn} \eeq

To go on, we use a method based on the Mellin-Barnes representation
for dealing with the sum over $l$ in the latter equation. In fact, for
$\Re z>1$ we have \cite{grad80b} \beq
\sum_{l=0}^\ii(2l+1)\rho^2K_{ir}^2(\rho\la_l)
=\frac{1}{4i\sqrt\pi}\int_{\Re z>1}
\frac{\Ga(z+ir)\Ga(z-ir)\Ga(z)}{\Ga(z+1/2)} \,\rho^{2-2z}\,f(z)\,dz
\:,\label{MBR} \eeq where \beq f(z)=\sum_{l=0}^\ii(2l+1)\la_l^{-2z}
\eeq is convergent for $\Re z>1$ and may be analytically continued to
the whole complex plane using standard methods (see for example
\cite{acto87-20-927,camp90-196-1}). In fact one has \beq
f(z)=2\sum_{n=0}^\ii \frac{(-1)^n\Ga(z+n)\at m^2+C-\frac1{4}\ct^n}
{\Ga(n+1)\Ga(z)} \,\ze_H(2z+2n-1,1/2) \:.\label{99} \eeq Here $\ze_H$
represents the Riemann-Hurwitz zeta-function. From the latter equation
it follows that $f(z)$ has only a simple pole at $z=1$ with residue
equal to $1$ and $f(0)=1/3-m^2-C$. Thus the integrand function in
Eq.~(\ref{MBR}) has simple poles at all the points $z=n$ ($n\leq1$)
and $z=n\pm ir$ ($n\leq0$).

Since we are mainly interested in global quantities, like the
partition function, we integrate Eq.~(\ref{bn}) over $\bar\M^3$,
paying attention to the fact that the integration over $\rho$ formally
gives rise to divergences. In order to regularize such an integral, we
introduce a horizon cutoff $\ep>0$ for small $\rho$. When possible, we
shall take the limit $\ep\to0$. In this way we have \beq \Tr
e^{-t\bar{L}_3}&=&\frac{A}{(4\pi)^{3/2}}\:
\int_{0}^\ii\,dr\,e^{-tr^2}\aq\phantom{\frac11} r\sinh\pi
r\:\X\cp\nn\\ &&\ap\hs\frac{1}{2\pi i}\int_{\Re z>1}
\frac{\Ga(z+ir)\Ga(z-ir)\Ga(z)}{z\Ga(z+1/2)} \,\ep^{-2z}\,f(z)\,dz\cq
\:,\label{4bh1} \eeq where the horizon area $A$ is equal to $4\pi
r_H^2$. The integrand function in Eq.~(\ref{4bh1}) has the same poles
as the function in Eq.~(\ref{MBR}), with the only exception that $z=0$
is a double pole. Now the advantage in that all the poles in the half
plane $\Re z<0$ give vanishing contributions to the integral in the
limit $\ep\to0$. So we get \beq \Tr
e^{-t\bar{L}_3}&=&\frac{A}{(4\pi)^{3/2}}\ag \frac{t^{-3/2}}{2\ep^2}
+\aq f(0)\ln\frac2\ep+\frac{f'(0)}2\cq t^{-1/2}\cp\nn\\
&&\ap\hs+\frac{f(0)}{8\sqrt\pi} +\frac{f(0)}{\sqrt\pi}\int_{0}^\ii
\aq\psi(ir)+\psi(-ir)\cq e^{-tr^2}\,dr\cg \label{4bh} \:,\eeq where
$\psi$ is the logarithmic derivative of the $\Ga$-function. In the
derivation, we made use of \beq
\lim_{\ep\to0}\frac{\ep^{-ix}}{x}=i\pi\de(x) \,.\eeq We note that in
the $t$ expansion of Eq.~(\ref{4bh}) appears a term independent of
$t$. Thus, we may write \beq \Tr e^{-t\bar{L}_3}=A\int_{0}^\ii
n(r)e^{-tr^2}\,dr+\frac{Af(0)}{64\pi^2} \label{c}\eeq where \beq
n(r)=\frac{r^2}{4\pi^2\ep^2} +\frac1{4\pi^2}\aq
f(0)\ln\frac2\ep+\frac{f'(0)}2\cq +\frac{f(0)}{8\pi^2}
\aq\psi(ir)+\psi(-ir)\cq \label{4bh2} \:.\eeq The last term in Eq.
(\ref{c}) may be omitted in the expression of the formal
$\zeta$-function, which reads \beq \ze(s|\bar{L}_3)=A\int_{0}^\ii n(r)
r^{-2s} dr \label{z3} \:.\eeq As a consequence, $\ze(-1/2|\bar{L}_3)$,
although formally  divergent, is "regular" at $s=-1/2$. This
"regularity" is due to the presence of the tachionic mass in the
conformally transformed operator $\bar{L}_3$ and its divergence is
related to the vacuum state of our quantization scheme in the
Schwarzschild coordinates, namely the Boulware vacuum. Formally, the
regularity of the quantity $\ze(-1/2|\bar{L}_3)$ is equivalent to the
absence of the conformal anomaly in the optical manifold.

The corresponding partition function may be evaluated making use of
Eq.~(\ref{lnZbeta}), with the replacement $\be\to\be/2$ due to the
redefinition of the Schwarzschild time in Eq.~(\ref{ro}). Finally we
have \beq \ln Z_\be &=&A\be j_\ep-\frac{\be}{4}\ze(-\fr12|\bar L_3)
+\frac{2\pi^2A}{45\ep^2\be^3} +\frac{A}{12\be}\aq
f(0)\ln\frac2\ep+\frac{f'(0)}2\cq \nn\\&&
-\frac{Af(0)}{64\pi^2}\ln\frac\be2 -\frac{Af(0)}{8\pi^2}
\int_0^\ii\ln\at1-e^{-\be r/2}\ct\, \aq\psi(ir)+\psi(-ir)\cq\,dr
\:,\label{bhf} \eeq where we have written the Jacobian contribution
due to the conformal transformation in the form $A\be j_\ep$. The
horizon divergences which appear in Eq.~(\ref{bhf}) are also present
in the statistical sum contribution to the free energy. The leading
one, due to the optical volume, is proportional to the horizon area
\cite{thoo85-256-727}, but in contrast to the Rindler space-time a
logarithmic divergence is also present, similar to the one found in
Ref.~\cite{solo94u-246,deal94u-27}.

As is well known, one needs a renormalization in order to remove the
vacuum divergences. We recall that these divergenges, as well as the
Jacobian conformal factor, being linear in $\be$, do not make
contribution to the entropy. However the situation here is complicated
by the presence of the horizon divergences, controlled by the cutoff
$\ep$. In the Schwarzschild space-time, it is known that the
renormalized stress-energy tensor is well defined at the horizon in
the Hartle-Hawking state \cite{scia81-30-327,brow85-31-2514}, which in
our formalism corresponds to the Hawking temperature $\be=\be_H$. This
means that the renormalized partition function has to be of the form
\beq \ln Z^R_\be &=&\frac{A\be}{90(8\pi)^2\ep^2}
\aq\at\frac{\be_H}{\be}\ct^4+3\cq -\frac{A\be}{3(8\pi)^2}\ln\ep
\aq\at\frac{\be_H}{\be}\ct^2+1\cq\nn\\ &&\hs+\frac{A}{12\be}\aq
f(0)\ln2+\frac{f'(0)}2\cq -\frac{Af(0)}{(8\pi)^2}\ln\frac\be2\nn\\
&&\hs\hs-\frac{Af(0)}{8\pi^2} \int_0^\ii\ln\at1-e^{-\be r/2}\ct\,
\aq\psi(ir)+\psi(-ir)\cq\,dr \:,\label{bhf2} \eeq where we have
introduced the Hawking temperature $\be_H=4\pi=8\pi MG$. The first
quantum corrections at temperature $T=1/\be$ to entropy, free and
internal energy immediately follows from Eq.~(\ref{bhf2}). In
particular, for the internal energy we have (the dots stay for finite
contributions at the horizon, which we do not write down because their
value depend on the approximation made) \beq
U_\be^R&=&\frac{A}{30(8\pi)^2\ep^2} \aq\at\frac{\be_H}{\be}\ct^4-1\cq
-\frac{A}{3(8\pi)^2}\ln\ep \aq\at\frac{\be_H}{\be}\ct^2-1\cq
\:\:+\dots \:,\eeq which has no divergences at $\be=\be_H$, while the
entropy \beq S_\be=\frac{8\pi^2A}{45\ep^2\be^3}
-\frac{A\ln\ep}{6\be}\:\:+\dots \label{S}\eeq also for $\be=\be_H$
contains the well known divergent term proportional to the horizon
area \cite{thoo85-256-727} and, according to Ref.~\cite{solo94u-246},
a logarithmic divergence too. Eq.~(\ref{S}) is vanishing in the
Boulware vacuum corresponding to $\be=\ii$.

We conclude with some remarks. In this paper the large mass limit of a
4-dimensional black hole has been investigated making use of conformal
transformation techniques, which allow one to work within the so
called optical manifold and some advantages have been achieved.  We
have succeded in obtaining a reasonable expression, valid only for
very large black hole mass and near the horizon. The leading
contribution gives rise to a divergence of the entropy  similar to the
one of the Rindler case, but other contributions in general are
present, leading to logarithmic divergences and finite parts, which,
however, depend on the initial approximation. With regard to this we
also would like to mention the results obtained in
Ref.~\cite{frol93-48-4545}, where the contributions to the black hole
entropy due to modes located inside and near the horizon have been
evaluated, using a new invariant statistical mechanical definition for
the black hole entropy. The finite contributions, namely the ones
indipendent on the horizon cutoff, are compatible with our results.

Finally few words about the horizon divergences. Physically, they may
be interpreted in terms of the infinite gravitational redshift
existing between the spatial infinity, where one measures the generic
equilibrium temperature and the horizon, which is classically
inaccessible for the Schwarzschild external observer. With regards to
the horizons divergences, we have argued that they are absent in the
internal energy at the Hawking temperature. However, they remain in
the entropy and in the other thermodynamical quantities, as soon as
one assumes the validity of the usual thermodynamical relations. A
possible way to deal with such divergences has been suggested in
Refs.~\cite{thoo85-256-727,frol93-48-4545,frol94u-36}, where it has
been argued that the quantum fluctuations at the horizon might provide
a natural cutoff. In particular, choosing the horizon cutoff of the
order of the Planck length ($\ep^2=G$), the leading "divergences",
evaluated at the Hawking temperature, turns out to be of the form of
the the "classical" Bekenstein-Hawking entropy. However one should
remark that other terms are present, giving contributions which
violate the area law. Alternatively one can try to relate the horizon
divergences to the ultaviolet divergences of quantum gravity, thus
arriving at the theory of superstring progagating in a curved
space-time \cite{suss94-50-2700}.

Finally, as far as  the Rindler space-time is concerned, this
space-time can be treated in a similar way. In our formalism, the
massless scalar field in Rindler space-time can be exactly solved,
because the optical spatial section turns out to be the hyperbolic
space $H^3$ and the harmonic analysis on such manifold is well known.
The computation of the quantum partition function is very similar to
the massive black hole space-time and we report it here for
comparison. The renormalized free energy may be chosen in the form
(here the proper acceleration $a=1$) \beq
F^R_\be=-\frac{A}{45(8\pi)^2\ep^2}\aq\at\frac{\be_U}{\be}\ct^4+3\cq
\:,\eeq where now the horizon area $A$ is infinite and $\be_U=2\pi
a^{-1}$ is the Unruh temperature. As a consequence, the entropy turns
out to be \beq S_\be=\frac{8\pi^2 A}{45\ep^2\be^3} \eeq and this
diverges for every finite $\be$, but is zero at zero temperature (the
Fulling-Rindler state), which is correct, since we are dealing with a
pure state. Furthermore, at $\be=\be_U$, corresponding to the
Minkowski vacuum, we have a divergent entropy, proportional to the
area, regardless of the fact that the Minkowski vacuum is a pure
state. This is also to be expected, since an uniformly accelerated
observer cannot observe the whole Minkowski space-time. Finally with
this renormalization prescription, the internal energy should read
\beq U^R_\be=\frac{A}{15(8\pi)^2\ep^2}
\aq\at\frac{\be_U}{\be}\ct^4-1\cq \eeq and this is vanishing and a
fortiori finite at $\be=\be_U$, as it should be. Furthermore, at
$\be=\ii$, namely in the Fulling-Rindler vacuum, it is in agreement
with the result obtained in Ref.~\cite{brow86-33-2840}.

\ack{We would like to thank A.A. Bytsenko. E. Elizalde and K. Kirsten
for discussions.} \end{document}